\documentclass{elsart4-1}
\usepackage[textwidth=460pt,textheight=690pt]{geometry}
\usepackage{graphicx}
\usepackage{psfrag}
\usepackage{amsmath}
\usepackage{amssymb}
\usepackage{bbold}
\usepackage{pstcol}
\usepackage{pst-node}
\usepackage{pst-plot}
\usepackage[numbers]{natbib}
\usepackage{hyperref}
\newcommand{\cal}{\mathcal}
\DeclareMathOperator{\Tr}{Tr}
\DeclareMathOperator{\Real}{Re}
\DeclareMathOperator{\Imag}{Im}
\DeclareMathOperator{\vol}{vol}

\newcommand\eqn[1]{(\ref{#1})}

\newcommand{\maxp}{cutoff}

\begin{document}
\makeatletter
\def\ps@copyright{\let\@mkboth\@gobbletwo
   \def\@oddhead{}%
   \let\@evenhead\@oddhead
   \def\@oddfoot{\small\slshape
      \def\@tempa{0}
       $^*$~Contribution to the proceedings of Strings 2004, Paris.\hfil\/%
       }
   \let\@evenfoot\@oddfoot
}
\makeatother
\begin{frontmatter}
\begin{flushright}
AEI-2004-104\\
hep-th/0412089
\end{flushright}
\vskip -4ex
\title{Holographic dynamics of unstable branes in AdS~$^*$}
\author{Marija Zamaklar and Kasper Peeters}
\ead{marija.zamaklar@aei.mpg.de}
\ead{kasper.peeters@aei.mpg.de}
\address{MPI/AEI f\"ur Gravitationsphysik, Am M\"uhlenberg 1,
  14476 Golm, GERMANY}
\begin{abstract}
The gauge theory dual to the decay of an unstable D-particle in AdS is
analysed in terms of coherent states. We discuss in detail how to
count the number of particles in the decay product. We find, in
agreement with the analysis in flat space, that the emission amplitude
is suppressed as the mass of the radiated particles increases.
\end{abstract}
\end{frontmatter}

\section{Introduction}

Tachyon condensation provides an interesting arena in which we can
improve our understanding of string theory in a dynamical set-up.
While the condensation of closed string tachyons, and the associated
decay of spacetimes, is still hampered by conceptual and technical
problems, a lot of progress has recently been made in understanding
the dynamics of open string tachyons.  Most of the analysis was
performed directly using boundary conformal field theory in flat
space, initiated by Sen's construction of the boundary states for
decaying D-branes~\cite{Sen:2002nu}, or by using the~$c=1$ matrix
model for the description of the decay of D-branes in 1+1 dimensional
string theory~\cite{McGreevy:2003kb}. In the present work we study the
problem of decaying branes in the set-up of the ``standard'' AdS/CFT
correspondence.

As was argued by Harvey et al.~\cite{Harvey:2000qu}, unstable D-branes
in string-theory are equivalents of ``sphalerons'': they are unstable
solutions located at a saddle point of the potential in string field
theory configuration space, at the top of a non-contractible
loop~\cite{Manton:1983nd}.  In the context of the AdS/CFT conjecture,
this correspondence between unstable D-branes and sphalerons in gauge
theory is in fact even more direct. By analysing the \emph{kinematical
properties} of these two systems, it has been argued by Drukker et
al.~\cite{Drukker:2000wx} that the unstable D-particles of string
theory are in precise correspondence with known sphaleron solutions of
the dual gauge theory.

We will study \emph{dynamical properties} of this correspondence. On
the gravity side we start with the results of Lambert et
al.~\cite{Lambert:2003zr} for the spectrum of decaying D-branes in
\emph{flat space}. To compare these results to those which we will
obtain in gauge theory, we ``embed'' the flat-space results in the AdS
space. A priori, there is no reason to expect that the flat space
results of the decay should be valid for branes in an AdS
background. However, since the D-particles in question are fully
localised in the bulk space, one expects that the flat space results
should carry over, at least when the radius of the AdS is large.

There are two properties of the spectrum of the decaying brane that we
want to compare with the dual gauge theory calculation. The first
property of the spectrum is constrained by the symmetries of the
system, and concerns emission amplitudes for the states on the leading
Regge trajectory. By slightly refining the calculation
of~\cite{Lambert:2003zr} we find \cite{Peeters:2004rd} that all
emission amplitudes for these states are zero. The same result is 
separately recovered on the gauge theory side by evaluating the number
operator for the corresponding dual composite operators.

More important is a second property of the spectrum, observed
in~\cite{Lambert:2003zr}, which reflects genuine dynamical features of
the decay. There is strong evidence~\cite{Sen:2002nu,Lambert:2003zr}
that the open strings decay \emph{fully} into closed string states,
i.e.~that there is no open string remnant left after the decay. This
conclusion is also supported by the matrix model calculations
of~\cite{McGreevy:2003kb}.  As shown in~\cite{Lambert:2003zr}, the
emission amplitudes are exponentially suppressed with the level of the
emitted string, at least for high levels (however, due to the
exponential growth of the available states, most of the energy of the
brane gets transferred into a high-density cloud of very massive
closed string states).

In the remainder of this report we focus on two issues in the dual
gauge theory on the boundary. The first issue is the construction of
the time-dependent gauge theory solution which is the analogue of
Sen's time-dependent boundary state in boundary conformal field
theory.  The second issue is how, given this time dependent solution,
one can reproduce the two properties of the spectrum of the decaying
particle mentioned above.

\section{D-particle $\leftrightarrow$ sphaleron correspondence: statics}

\begin{figure}[t]
\hskip 3em\hbox{\vbox{\psfrag{E}{$\!\!\!\!E_{\text{p}}$}
\includegraphics*[width=.25\textwidth]{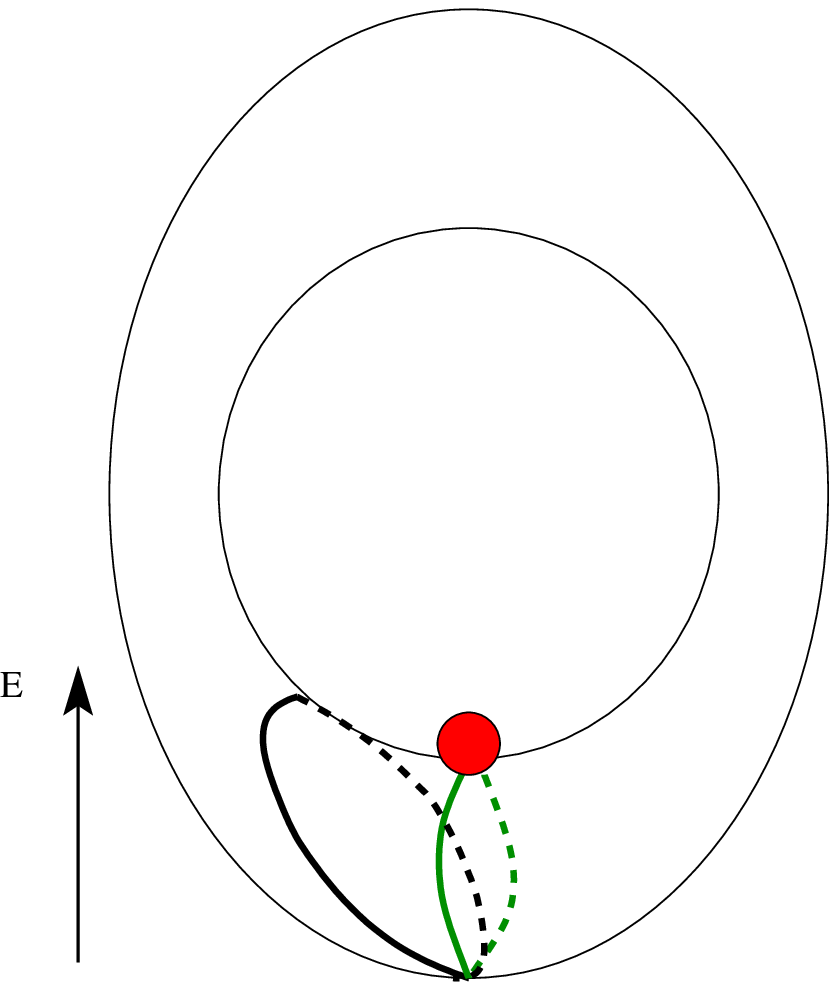}}\hskip-.6\textwidth
\vbox{\psfrag{S(r)}{\small $S_{\text{inst}}(r)$} \psfrag{r=a,
f(a)=1/2}{\small \vbox{\hbox{$r=a$}\hbox{$f=\frac{1}{2}$}}}
\psfrag{r=0}{\small $r=0$} \psfrag{s}{\rnode{sphal_t}{\phantom{s}}}
\psfrag{r}{\small $r \rightarrow$}
\includegraphics*[width=.5\textwidth]{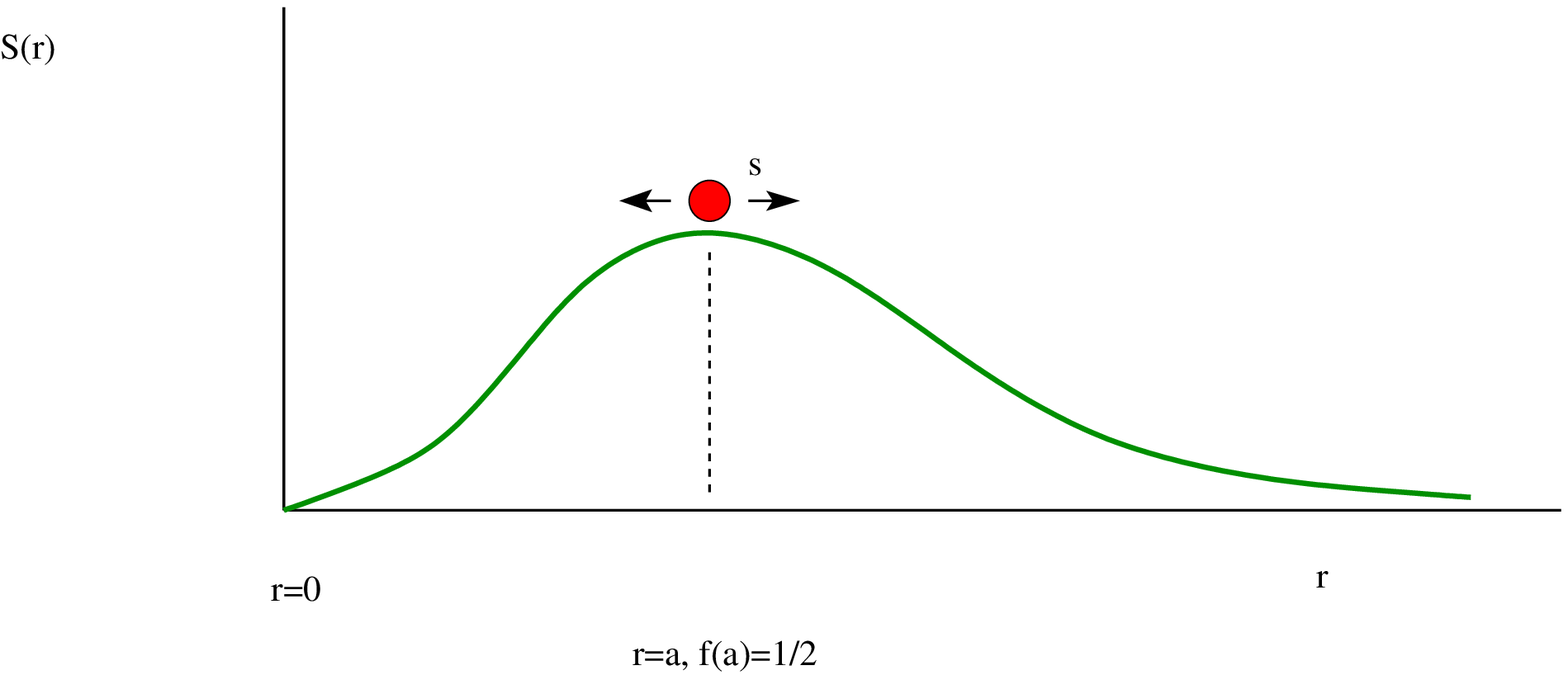}}}
\caption{The picture on the left shows in a schematic way the
  existence of a non-contractible loop in configuration space, as well
  as the presence of the sphaleron (red dot) at the saddle point. The
  picture on the right shows the action density $S(r)$ of the
  instanton in the Euclidean theory, together with the special
  configuration at $r=a$ which is used to construct the sphaleronic
  particle in the Lorentzian theory.}
\label{f:configspace}
\end{figure}

Before addressing the dynamical properties of the correspondence, let
us first briefly revise its basic static
properties~\cite{Drukker:2000wx}.  Gauge theory
sphalerons~\cite{Manton:1983nd} are static solutions of the equations
of motion, associated to saddle points whose existence is guaranteed
by the existence of a non-contractible loop in (compact) configuration
space.  Whereas the sphaleron solution on~$\mathbb{R}^4$ in
Yang-Mills-Higgs theory, found by Klinkhamer and Manton
\cite{Klinkhamer:1984di}, is very complicated and not known
analytically, the situation is much simpler on $S^3\times\mathbb{R}$
in pure Yang-Mills theory.  To construct the sphaleronic gauge
configuration in the SU(2) gauge theory, one starts from the
instanton solution on $\mathbb{R}^4$,
\begin{equation}
\label{e:ansatz}
A_{\mu} = f(r) (\partial_{\mu}U) U^{\dagger}, \quad \quad
U=\frac{x^{\mu}\sigma_{\mu}}{r}, 
\quad \quad r^2 = x_0^2+x_i^2  \,,
\end{equation} 
where $f= r^2/(r^2 + a^2)$. This function interpolates between two
pure gauge configurations (i.e.~the two vacua) $f(r=0)=0$ and
$f(r=\infty)=1$. When $f(r)=1/2$, the system is at the top of the
potential barrier, see figure~\ref{f:configspace}. By taking $f=1/2$
everywhere one gets a singular solution to the equation of motion on
$\mathbb{R}^4$, which is the so-called ``meron''. The $f=1/2$ solution
is, however, also a solution on $S^3\times \mathbb{R}$, since this
manifold can be conformally mapped to $\mathbb{R}^4$ and Yang-Mills
theory in four dimensions is conformally invariant. The solution
obtained in this way is the Euclidean version of the ``sphaleron'',
and is non-singular.
The Lorentzian version is the same, since the time component of the
potential of the sphaleron is zero. The solution is completely
time-independent and has infinite action, corresponding to a
sphaleronic particle which is sitting at the top of the potential.

As far as a generalisation of the previous construction to SU($N$)
gauge theory is concerned, the general sphaleron configuration is not
known. However, an interesting special configuration has been given
in~\cite{Drukker:2000wx}. It is obtained by replacing the Pauli
matrices in~\eqn{e:ansatz} with Clifford algebra generators according
to
\begin{equation}
\label{e:sunsu2}
\sigma_\mu \rightarrow \gamma_\mu 
 = \begin{pmatrix}
   \sigma_\mu & 0 & \cdots & 0 \\
   0          & \sigma_\mu & \cdots & 0 \\
   \vdots     & \vdots & \ddots & 0 \\
   0          & 0      & \cdots & \sigma_\mu 
   \end{pmatrix}\, .
\end{equation}
The mass of this sphaleronic particle is $k$~times the mass of the
SU(2) particle (where $k$ is the number of sigma matrices
in~\eqn{e:sunsu2} and $2k < N$). It was also shown that the number
unstable modes is increased from one (for SU(2)) to $k^2$.

It has been argued by Drukker et al.~\cite{Drukker:2000wx} that the
(non-supersymmetric) sphaleronic saddle points in the gauge theory are
preserved as the 't~Hooft coupling is increased, despite the fact that
the precise form of the potential receives quantum corrections. The
main reason for this is that these sphaleronic saddle points are
linked to the underlying non-contractible loops in configuration
space. Furthermore, they are linked to the (supersymmetric) instanton
configuration which is present both at strong and weak coupling. Thus
the sphaleronic particle in the Yang-Mills theory on $S^3 \times R$
has, at weak coupling, been conjectured to be dual to the unstable
D-particle in the AdS.

A number of arguments has been given~\cite{Drukker:2000wx} in support
of this correspondence.  Firstly, both D-particles and sphaleronic
particles are static with respect to the global AdS time. Secondly,
since the D-particle is located at the origin of the AdS space (in
global coordinates), it is ``projected'' in a homogeneous fashion to
the boundary, in agreement with the fact that the sphaleronic particle
is homogeneously spread over the $S^3$. Thirdly, the D-particle in the
bulk is a source for the gravitational and dilaton field (while it
does not source the RR forms), which is in agreement with the
(non)vanishing expectation values of the dual gauge
operators. Finally, in the case of the more general
sphaleron~\eqn{e:sunsu2}, the number of unstable modes on both sides
agrees.

\section{D-particle $\leftrightarrow$ sphaleron correspondence: dynamics}

To study the dynamics of the decaying D-particles from the gauge
theory perspective let us, as a first step, construct the time
dependent gauge configuration describing the sphaleron decay. We
restrict to the decay modes which preserve spherical symmetry by
making the following ansatz,
\begin{equation}
\label{ansatz}
A = f(t)\, \Sigma^i \sigma_i\,,
\end{equation}
where $\Sigma^i$ are the three left-invariant one-forms.  To deduce
what is the unknown function $f(t)$ we plug the ansatz into the action
and derive the action for this function. The value of the action for
our ansatz is
\begin{figure}[t]
\begin{center}
\includegraphics*[width=.5\textwidth]{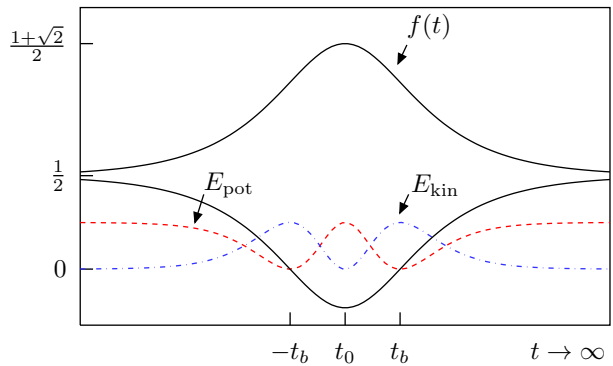}
\caption{The functions $f_\pm(t)$ of the decaying sphaleron on $S^3$ as
  given in~\eqn{e:fsol}, together with the kinetic and potential
  energy (with normalisation as given in~\eqn{e:theenergy} and $R=1$).}
\label{f:foft}
\end{center}
\end{figure}%
\begin{equation}
\label{e:reducedYM}
S = -\frac{1}{4 g_{\text{YM}}^2} \int\!{\rm d}t{\rm d}\Omega\, F_{\mu\nu}F^{\mu\nu}
  =  \frac{24\, \vol(S^3)}{4\, g_{\text{YM}}^2} \int\! \frac{{\rm d}t}{R} 
       \left(\frac{R^2}{2}\dot{f}^2 - 2 f^2 (1-f)^2\right) \, ,
\end{equation}
where $\vol(S^3)\equiv 2\pi^2$ denotes the volume of the unit sphere
and~$R$ is the radius of $S^3$. The equation of motion for the
function~$f$ is
\begin{equation}
\label{single}
R^2\,\ddot{f} + 4f(1-f)(1-2f) = 0 \, .
\end{equation}
When integrated once, this equation yields a conserved quantity,
namely the energy (i.e.~the component $T_{00}=48 \vol(S^3) E$)
\begin{equation}
\label{e:theenergy}
E = R^2\,\dot{f}^2 + 4 f^2 (1-f)^2 \, 
\end{equation}
which is simplest to integrate analytically
for~$E=\tfrac{1}{4}$. There are two solutions, corresponding to the
fact that the sphaleron can roll down on either side of the potential,
to the vacua with Chern-Simons number one and zero respectively. The
final result reads (see figure~\ref{f:foft})
\begin{equation}
\label{e:fsol}
f_\pm(t) = \frac{1}{2}\left(\frac{\pm\sqrt{2}}{\cosh\left(\frac{\sqrt{2}}{R} (t-t_0)\right)}
+ 1 \right)\, .
\end{equation}
This solution describes a configuration that starts from the potential
maximum at $t=-\infty$ (with zero velocity and acceleration), rolls
down the hill and up the other side, where it arrives at
$t=t_0$.\footnote{After we had derived this solution, we learned that
it has been obtained before~\cite{Gibbons:1994pq} albeit in a
different context.}

\begin{figure}[t]
\begin{center}
\psfrag{tb}{\raise -1ex\hbox{$t_b$}}
\psfrag{decay}{\!\!\!``decayed D0''}
\includegraphics*[width=.3\textwidth]{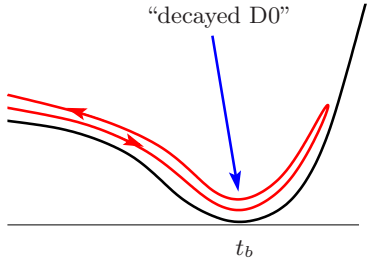}
\end{center}
\caption{The evolution of the sphaleron. As it rolls down, it reaches
  a point where all potential energy has been converted to kinetic
  energy. This is what we will call the ``decayed D-particle'',
  despite the fact that the decay products will eventually come back
  as fine-tuned radiation to ``re-build'' the D-particle.}
\label{f:rolling}
\end{figure}

The periodicity of the whole process is natural from the AdS
perspective. Since AdS effectively acts as a box, the cloud of
outgoing radiation is refocused to the origin of the space, where it
arrives as fine-tuned radiation and ``re-builds'' the D-particle. In
this sense the D-particle never decays, since there is no real
dissipation of the energy in the system. However, in the limit of
large AdS radius, our flat-space intuition should (at least
approximately) hold. A natural point in time, which should be
associated to the decayed brane, is the point where the sphaleron has
rolled down to the the bottom of the potential, i.e.~when all
potential energy has been converted to kinetic energy (see
figure~\ref{f:rolling}).

Near the bottom the solution is
\begin{equation}
\label{e:Afull}
A_{\mu} = \tilde{f}(t)\, U^\dagger (\partial_\mu U)  \, , \quad
 \tilde{f} = f-1 \, .
\end{equation}
with $\tilde{f}\approx 0$, which means that the derivative part of the
field strength, rather than the non-linear (commutator) part, is
dominant. The solution becomes a solution of the free Yang-Mills
equations of motion on~$S^3 \times \mathbb{R}$ (written in the
radiation gauge: $A_0 = \nabla_i A^i=0$),
\begin{equation}
\label{e:linearisedEOM}
\Big( -\partial_t^2 + \frac{1}{R^2} \big(\nabla_{S^3}^2 - 2\big) \Big) A^{\text{lin.}}_i =
0\, .
\end{equation}
Indeed, one can easily see that as~$t \rightarrow t_{\text{bottom}}$
the solution~\eqn{e:Afull} with $f$ given by~\eqn{e:fsol} is
very well approximated by the following solution of the linearised
equation of motion~\eqn{e:linearisedEOM}:
\begin{equation}
\label{e:Anearbottom}
A_i^{\text{lin.}} = 
-\frac{1}{4}\sin\left(\frac{2(t-t_{\text{bottom}})}{R}\right) \,
U^\dagger (\partial_i U) \,.
\end{equation}
Hence near the bottom of the valley, one can think about the
Yang-Mills configuration as dual to a coherent state of
non-interacting closed string states which are the product of the
D-particle decay.  Our goal will then be to determine the numbers of
various (gravity) ``particles'' in this final coherent state. What we
precisely mean by this will be explained in the next section. Let us
first construct this coherent state.

The fact that our solution abelianises near the bottom of the
potential allows us to apply the standard machinery to write down the
coherent state. By expanding the classical, free Yang-Mills gauge
potential in terms of spherical vector harmonics, one can read off the
amplitudes for different modes, and write a coherent state as
\begin{equation}
\label{coherent}
|c\rangle = {\mathcal C}\, \exp\left( g_{\text{YM}}^{-2} \sum_{J,M,y} 
   \Tr\big(A_{JMy}\, \hat{a}^{\dagger}_{JMy}\big) \right) |0\rangle\,,
\end{equation}
where $A_{JMy}$ are the coefficients appearing in the Fourier
decomposition of the classical sphal\-eron configuration and the
normalisation factor~${\mathcal C}$ is chosen such that $|c\rangle$ is
of unit norm.

For this to be a legitimate state in the Hilbert space, one has to
make sure that it satisfies all constraints. It is easy to see that
creation operators in~\eqn{coherent} lead to physical excitations in
the free theory. However, once $g_{\text{YM}}$ is turned on, Gauss' law
implies that only singlets can be excited. This means that all
non-singlet states in~\eqn{coherent} have to be projected out.  In
practice, however, we will neither write this projector nor construct
the projected state explicitly. This is because our calculations
always involve projections of the coherent state onto states which
themselves are color singlets. Therefore the singlet projection is
imposed implicitly throughout.

\section{Particles in the AdS/CFT correspondence}

In the AdS/CFT correspondence we have a relation between string states
in the bulk and operators in the boundary. These operators are, via
the operator--state mapping, interpreted to create ``particles'' in
the bulk theory at a particular point on the boundary. That is, one
needs to solve for the wave equation of the dual field in the bulk in
the presence of a delta source inserted at the boundary.  This means
that the states created in the bulk are not eigen-momentum states, an
attribute which one usually associates to the notion of a particle in
field theories.  However, since the AdS/CFT correspondence is
formulated in position space rather than momentum space, these
definitions are natural in this context.  On the other hand, our
string calculation in~\cite{Peeters:2004rd} is a flat space
calculation, and for us it will be more natural to use the standard
notion of particles in the bulk as angular momentum
eigenstates. Therefore, we will first have to construct boundary
operators that are dual to bulk angular momentum eigenstates.

The operator--state correspondence is usually discussed in the context
of radial quantisation of conformal field theories (see
e.g.~\cite{Fubini:1973mf} for a discussion in a four-dimensional
context). One first Wick rotates $\mathbb{R}\times \mathbb{R}^3$ to
the Euclidean regime and then performs a conformal transformation such
that the origin of $\mathbb{R}^4$ corresponds to $t=-\infty$ in the
original frame.  Operators inserted at the origin are then in
one-to-one correspondence with states in the Hilbert space. The entire
procedure can, however, be formulated without doing the conformal
rescaling, which is more natural in our setup since, as we have
discussed before, the gauge field configuration on $\mathbb{R}\times
S^3$ is non-singular while the one on $\mathbb{R}^4$ is singular.

The state corresponding to an operator with conformal weight~$w$ is
obtained by multiplying with the appropriate exponential of Euclidean
time and taking the limit $\tau\rightarrow-\infty$ (keeping only the
regular part): 
\begin{equation}
\label{e:Ostatedef}
|\hat{O}^{(m)}_{\text{weight-}w}\rangle = \lim_{\tau\rightarrow-\infty} 
  \,\Big\{e^{-w\tau}
  \hat{O}^{(m)}_{\text{weight-}w}(\tau)\Big\}
  \big|0\big\rangle
\equiv \hat{O}^{\dagger(m)}_{\text{weight-}w} |0\rangle\, .
\end{equation}
The last expression shows the shorthand notation that we will use in
order not to clutter expressions unnecessarily.  The hermitian
conjugate of an operator is given by
\begin{equation}
\Big(\hat{O}(\tau)\Big)^\dagger
 = \hat{O}^\dagger(-\tau)\, .
\end{equation}
This procedure mimics the operator--state mapping on $\mathbb{R}^4$
but avoids technical problems related to solutions which become
singular after the conformal transformation.

The operators which we use in~\eqn{e:Ostatedef} are independent of the
angular coordinates on the sphere, i.e.~they are obtained from the
position dependent operators as follows
\begin{equation}
\label{e:pola}
\hat{O}_{w}^{(m)}(\tau) = K^{(m)}_{w} \int_{S^3}\! {\rm d}\Omega\,
\, \hat{O}_{w}^{\mu_1...\mu_s}(\tau,\phi_i)\, Y_{\mu_1...\mu_s}^{(m)}(\phi_i) \, .
\end{equation}
Here $Y^{(m)}$ denote the lowest lying tensor spherical harmonics for
a given spin~$s$. The index~$m$ labels the degeneracy of such
harmonics. The normalisation constants~$K^{(m)}_w$ are chosen such
that the states constructed using~\eqn{e:Ostatedef} are of unit norm.
Note that the multiplication with the time dependent exponent
in~\eqn{e:Ostatedef} selects out composite operators of the required
conformal dimension, but when one expresses these operators in terms
of elementary creation and annihilation operators, one explicitly sees
that different operators~$\hat{O}$ are not orthogonal.  It is only
after the integration~\eqn{e:pola} that one obtains a set of
orthogonal states. 

There are many subtleties related to the fact that operators $\hat{O}$
are composite operators rather than elementary gauge
operator. Firstly, the multi-particle states cannot simply be obtained
by acting repeatedly with the~$\hat{O}^\dagger$ operators on the
vacuum. States generated in this way are \emph{not orthogonal}, not
even in the $N\rightarrow\infty$ limit when the number of operators
becomes large as well. Starting from the naive states
$(\hat{O}^\dagger)^n|0\rangle$ one has to subtract terms in order to
achieve orthogonality. For the same reason, there is no simple number
operator which can be used to count the number of composite
excitations in a given state. It is true that
\begin{equation}
\label{e:OdaggerO}
{} [\hat{O}, \hat{O}^\dagger ] = 1 + {\mathcal O}(N^{-2}) \,,
\end{equation}
and one might expect that this leads to a well-defined number
operator~$\hat{O}^\dagger \hat{O}$. However, the coefficients that
multiply the~$1/N^2$ corrections in~\eqn{e:OdaggerO} are operators,
not c-numbers. As a consequence, the strength of the~$1/N^2$
corrections depends on the state in which the number operator is
evaluated,
\begin{equation}
\label{noco}
\langle n | \hat{O}^\dagger \hat{O} | n\rangle = n + \sum_i \frac{c_i(n)}{N^{2i}}\,.
\end{equation}
The numbers $c_i(n)$ can become arbitrarily large
when~$n\rightarrow\infty$. Since the coherent state contains such
highly excited states, the operator $\hat{O}^\dagger \hat{O}$ cannot
be used as a number operator, not even in the $N\rightarrow\infty$
limit.\footnote{An proper number operator for composite particles,
which produces the exact occupation number rather than an expression
which is only correct up to $N^{-2}$ corrections, has been constructed
by~\cite{Brittin:1980ev}. However, their operator is very
complicated and difficult to handle in practice. We prefer to follow a
different route here.} We will encounter an explicit manifestation of
these problems in the next section, when we start counting particles
in the coherent state, and then on a concrete example we will
illustrate how one can deal with them.

Let us end this section with a comment on alternatives to the coherent
state~\eqn{coherent}. From the point of view of the dual string
theory, it might seem more natural to construct a coherent state using
the composite operators~$\hat{O}^\dagger_J$ in the exponent, rather
than the elementary ones~$\hat{a}^\dagger$. After all, the~$\hat{O}_J$
correspond to elementary string excitations. However, a state of
the form
\begin{equation}
|\tilde c\rangle = \tilde{{\mathcal C}} \exp\Big(
  \sum_i O^{\text{class.}}_i\,\hat{O}^\dagger_i
\Big)|0\rangle
\end{equation}
is not a coherent state in the standard sense since the 
expectation value of an operator in this coherent state does not equal
the classical value of that operator,
\begin{equation}
\big\langle \tilde c\big|\, \hat{O}_i\, \big|\tilde c\big\rangle \not= O^{\text{class.}}_i\,,
\end{equation}
not even up to $1/N$ corrections. The reason for this is essentially
given in equation~\eqn{noco}, with~$|n \rangle$ now being given
by~$|n\rangle = \big(\hat{O}^\dagger_i\big)^n\,|0 \rangle$.  This is
our prime motivation to use~\eqn{coherent} as the sphaleron coherent
state.

\section{Particle counting}

Starting from the coherent state~\eqn{coherent} we now want to extract
information from it about particle numbers in the decay product. By
particle counting, we mean counting of the states constructed in the
previous section.

Due to the problems explained around~\eqn{e:OdaggerO}, one cannot use
the ``standard'' number operator $\hat{O}^\dagger\hat{O}$. Instead we
will simply decompose the coherent state on the basis of
multi-particle states. Subsequently we will, using these
probabilities, calculate the average energies and particle numbers.
The probability of finding a multi-particle state consisting of $p_1$
particles of type $O_{J_1}$, $p_2$ particles of type $O_{J_2}$ etc.,
is given by
\begin{equation}
\label{expectations}
{\mathcal P}(p_1;p_2;\ldots;p_M) :=
     \frac{\left|\, \Big\langle 
     (\hat{O}_{J_1})^{p_1} \ldots (\hat{O}_{J_M})^{p_M}\,
\Big|\,c\Big\rangle \, \right|^2}{\Big\langle \big( \hat{O}_{J_1} \big)^{p_1} \ldots 
            \big( \hat{O}_{J_M} \big)^{p_M}\, \Big|\,
            \big( \hat{O}_{J_1} \big)^{p_1} \ldots 
            \big( \hat{O}_{J_M} \big)^{p_M} \Big\rangle\,\big\langle
     c\big|c\big\rangle}\,.
\end{equation}
For this to work it is of course crucial that the basis of
multi-particle states is constructed to be orthogonal. By definition,
the average number of particles of the type~$\hat{O}_{J_i}$ present in
the coherent state is now given by
\begin{align}
\label{e:numbers}
N(J_i) &:= 
  \sum_{p_1=0}^\infty \cdots \sum_{p_M=0}^\infty
     p_i\, {\mathcal P} (p_1;p_2;\ldots;p_M)\,. 
\intertext{The energy stored in these particles, as measured with
respect to the global time in the bulk, is given by the conformal
dimension of the corresponding operators. Therefore, the total energy
  is given by the expression}
\label{e:energies}
E(J_i) &:= 
  \sum_{p_1=0}^\infty \cdots \sum_{p_M=0}^\infty
     \Delta_{J_i}\, p_i\, {\mathcal P} (p_1;p_2;\ldots;p_M)  \, , 
\end{align}
where $\Delta_{J_i}$ is the conformal dimension of the operator
$\hat{O}_{J_i}$.  For a generic operator, the calculation of the
numerators in~\eqn{expectations} reduces to evaluating the classical
expression of the (abelianised) operator using the positive frequency
part of the decayed solution.  Hence, by considering only the
numerators in~\eqn{expectations} we can deduce which particles are
\emph{absent} from the decay spectrum. In particular one can easily
deduce that expectation values of the operators dual to the graviton,
NS-NS two form and all twist two operators are zero.\footnote{Note
that the expression which vanishes is the energy momentum tensor
evaluated on the positive frequency part of the solution:~$|\langle 0
|\hat{T}_{\mu\nu}|c \rangle|^2 =
|T_{\mu\nu}(A^+_{\text{coherent}})|^2=0$. On the other hand, the
classical expression for the energy momentum tensor of the full
configuration is non-zero:~$T_{\mu\nu}(A^+ + A^-) \neq 0$.}  By
slightly refining the calculation of~\cite{Lambert:2003zr} we have
found that all emission amplitudes for these states are zero in string
theory as well~\cite{Peeters:2004rd}.  The absence of the
gravitational radiation is not surprising, since the decay is
spherically symmetric. We also believe that absence of the other
states is dictated by some underlying symmetry arguments.

Thus, to explore the genuine symmetry aspects of the decay we need
to concentrate on the states for which~\eqn{expectations} does not
vanish. The main technical problem arises when evaluating the denominators
of~\eqn{expectations}. To illustrate this, let us consider a
``simplified'' model, based on a non-abelian scalar field. This model
exhibits all of the technical subtleties associated to the
determination of the decay products. The crucial ingredients of the
vector coherent state, namely that it is constructed from the
lowest-lying spherical harmonics and that it depends
non-perturbatively on the coupling constant, are preserved by this toy
model. It, however, avoids the inessential technical complications
associated to the evaluation of tensor spherical harmonics in the
numerators of~\eqn{expectations}.

The coherent state for a given classical configuration in this
non-abelian scalar theory is given by
\begin{equation}
\label{e:simplecoh}
|c\rangle = {\mathcal C} \exp\left( \frac{1}{g_{\text{YM}}^2} \Tr\big( a\,
\hat{a}^\dagger\big) \right) |0\rangle\,, \qquad {\mathcal C} = \exp\left(
-\frac{1}{g_{\text{YM}}^2} \Tr\big( a^\dagger a\big)\right)\,.
\end{equation}
This mimics the construction~\eqn{coherent}. The unit normalised (at
leading order in $1/N$ expansion), single-trace operators which create
particles in the out vacuum are
\begin{equation}
\label{sample}
\hat{O}_J^\dagger = \frac{1}{\sqrt{\strut J (g_{YM}^2 N)^J}} \Tr\big( (\hat{a}^\dagger)^J \big) \, .
\end{equation}
These operators are coordinate independent operators, obtained using a
procedure similar to~\eqn{e:pola}.

With the above normalisation of the operator, the numerators and hence
probabilities in~\eqn{expectations} depend on the Yang-Mills coupling
in a non-perturbative fashion,
\begin{equation}
\label{argu}
\Big|\langle 0 | \big(\hat{O}_J\big)^p |c \rangle\Big|^2 = {\mathcal C}^2
				 \left| \frac{ \Tr\big((a^+)^J\big)}{\sqrt{\strut J (
				 g_{YM}^2 N)^J}} \right|^{2p} \, \equiv \frac{{\mathcal C}^2}{
				 J^p} \, \left({\frac{\eta_J^{2}}{\lambda^{J}}} \right)^p
				 \, ,
\end{equation}
(where the last equality defines $\eta_J$; note that it is of the
order~$N$ for the configuration~\eqn{e:sunsu2} and generically scales
as the number of D-particles).  This reflects the fact that our
original sphaleron configuration is a non-perturbative solution of the
equations of motion. Note also that the only way in which the
coupling~$\lambda$ appears in~\eqn{e:numbers} and~\eqn{e:energies} is
through the combination~$\eta_J^2/ \lambda^J$.

The complicated part of the calculation of the average particle
numbers and energies is the computation of the norms for the states
with an arbitrary number of particles.  The norm of the state with
$p$~identical particles can be written as 
\begin{equation}
\label{expansion}
\begin{aligned}
 \big\langle (\hat{O}_J)^p\, (\hat{O}_J^\dagger)^p \big\rangle &= 
   p!\, \big\langle (\hat{O}_J)\, (\hat{O}_J^\dagger)\big\rangle^p 
 + \binom{p}{2}^2 \big\langle
   (\hat{O}_J)^2\,(\hat{O}_J^{\dagger})^2\big \rangle_{\text{connected}} 
   (p-2)! \big\langle (\hat{O}_J)\, (\hat{O}_J^\dagger) \big\rangle^{(p-2)} \\ 
& +\binom{p}{3}^2 \langle  \hat{O}_J^3  \hat{O}_J^{\dagger 3} \, \rangle_{\text{connected}} (p-3)! \langle
\hat{O}_J \hat{O}_J^\dagger \rangle^{(p-3)} \\[1ex]
&+ \binom{p}{2}^2 \binom{p-2}{2}^2  \big\langle  (\hat{O}_J)^2
   (\hat{O}_J^{\dagger})^2  \big\rangle_{\text{connected}}^2  \frac{(p-4)!}{2!} \langle
\hat{O}_J \hat{O}_J^\dagger \rangle^{(p-4)} + ...
\end{aligned}
\end{equation}
The first term is at a leading order independent of $1/N$, the second
is suppressed as~$1/N^2$, the last two terms both scale as~$1/N^4$, and
so on.  A similar but more complicated expansion can be written for
states involving more than one type of particle.

Naively, one might expect that in the large-$N$ limit, all but the
leading term~$p!$ in this expansion can be omitted. However, this
would produce an exponential dependence on the expectation values for
the operators $\hat{O}_J$ in formula~\eqn{e:numbers}.  Since the
arguments of the exponent~\eqn{argu} \emph{increase} with conformal
dimension~$J$, one would conclude that the number of particles
produced during the decay \emph{increases} with the mass of the
particle.  It is easy to see that this kind of truncation
of~(\ref{expansion}) does not make sense in the case of the
\emph{non-perturbative} coherent state~\eqn{e:simplecoh}, as it would
actually produce probabilities~\eqn{expectations} which are larger
than one. The point is that since the numerator~\eqn{argu} is very
large, the maximal probabilities are attained for large
values~$p^{\text{max}}$ of~$p$. Moreover,~$p^{\text{max}}$ grows
with~$N$, hence in the large-$N$ limit the sub-leading terms
in~\eqn{expansion} become more and more relevant, and are actually
\emph{comparable} to the leading term.

In trying to estimate how fast the norms (\ref{expansion}) have to
grow with~$p$, one can see that even an exponential growth of the
norms, say as~$p!\, \gamma^p$ ($\gamma=\text{const}.$), does not lead
to reasonable results.  Namely, if we consider the
expression~$\sum_{p} {\mathcal P}(J,p)$, which has to be smaller than one,
and assume exponential growth of norms, we would find that this sum
behaves as
\begin{equation}
\label{e:pos-prob}
\sum_{p=0}^\infty {\mathcal P}(J,p) = {\mathcal C}^2 \sum_{p=0}^\infty
\frac{1}{p!} \left(\frac{\eta_J^2}{\lambda^J \gamma}\right)^p =
\exp\left(\frac{\eta_J^2}{\lambda^J \gamma}\right) \exp\left(- \frac{N}{\lambda}
\Tr(a^\dagger a)\right)\,.
\end{equation}
Hence we see that even when $N\rightarrow\infty$ (while keeping
$\lambda$ arbitrary but smaller than one) the result will always be
larger than~1 for some value of~$J$. Since the calculation of the
average number of particles requires a summation over all~$J$, we
conclude that we cannot assume this behavior of the norms.\footnote{Note
that if we would have had a perturbative coherent state instead of a
\emph{non-perturbative} one, the classical expectation values~$a$ in
(\ref{e:simplecoh}) would be of the form $a = g_{YM} \eta$, with
$\eta$ a number independent of the coupling constant.  Hence formula
(\ref{e:pos-prob}) would be replaced with
\begin{equation}
\label{e:pert-prob}
\sum_{p=0}^\infty {\mathcal P}(J,p) = {\mathcal C}^2 \sum_{p=0}^\infty
\frac{1}{p!} \left(\frac{\eta_J^2}{N^J \gamma}\right)^p =
\exp\left(\frac{\eta_J^2}{N^J \gamma}\right) \exp\left(- \Tr(a^\dagger a)\right)\,.
\end{equation}
We now see that a truncation to the first term in~\eqn{expansion}
(i.e.~setting $\gamma=1$) produces reasonable results for the
probabilities~\eqn{expectations}.}

The situation which we face here is similar in spirit to the
double-scaling BMN limit. As observed in~\cite{Kristjansen:2002bb}
and~\cite{Constable:2002vq}, in the limit $N \sim J^2
\rightarrow\infty$ correlators in general receive contributions from
non-planar graphs of all genera.  In this case, a new expansion
parameter~$J^2/N$ appears. In our case, $N\rightarrow\infty$ as well,
but now the additional parameter which becomes large is the value of
the~$p_i$ for which the sum~\eqn{e:energies} has its maximum term.  It
would be interesting to understand whether our system also exhibits a
double-scaling limit in which some ratio of powers of~$p$ and~$N$ is
kept fixed.

\section{Calculation of norms and numerical results}
\label{s:U4results}
In order to determine the correct values of the norms of the states,
it is useful to write the norms of multi-particle states in terms of
correlators of a complex matrix model,
\begin{multline}
\label{e:mcintegral}
\big\langle 0 \big| \Big[ \big( \hat{O}_{J_1} \big)^{p_1} \ldots 
            \big( \hat{O}_{J_n} \big)^{p_n}\Big] \, \Big[
            \big( \hat{O}^\dagger_{J_n} \big)^{p_n} \ldots 
            \big( \hat{O}^\dagger_{J_1} \big)^{p_1}\Big] \, \big|0
            \big\rangle\\[1ex]
 = \int\!{\rm d}A{\rm d}\bar A\, \Big[\big( {O}_{J_1} \big)^{p_1} \ldots 
            \big( {O}_{J_n} \big)^{p_n}\Big] \, \Big[
            \big( {O}^\dagger_{J_n} \big)^{p_n} \ldots 
            \big( {O}^\dagger_{J_1} \big)^{p_1}\Big] 
\exp\Big( - \Tr( A^\dagger A ) \Big)\,.
\end{multline}
The measure used here is simply a separate integral over the real and
imaginary parts of the complex matrix $A$, normalised to give unit
result when all~$p_i$ in the expression above are zero,
\begin{equation}
\int\!{\rm d}A{\rm d}\bar A = \pi^{-N}\prod_{a,b=1}^N {\rm d}(\Real
  A_{ab})\,\,\, {\rm d}(\Imag A_{ab})\,.
\end{equation}
This approach has been used
by~\cite{Kristjansen:2002bb,Beisert:2002bb} in order to compute
several special cases of~\eqn{e:mcintegral} analytically. It is still
an open problem to extend those exact results to the entire class of
correlators, in particular to general situations for
which~$p_i>2$. Because we will need these very general correlators, we
have decided to use an alternative approach, in which the integral is
evaluated using Monte-Carlo methods. This provides us with a
technically straightforward way to extract the norms for arbitrary
operator insertions, even for very large~$p_i$. Our results will, for
this reason, of course be restricted to a fixed value for~$N$ and
computer resources put a practical limit on the maximum value that can
be handled (we will take $N=4$). Nevertheless, we will see that
interesting results can be obtained this way.

In the U(4) theory there are only two operators which create
physical states (using only the creation operator for the lowest-lying
spherical harmonics). These are~$\Tr\big((a^\dagger)^2\big)$ and
$\Tr\big((a^\dagger)^4\big)$.\footnote{The restriction to the
zero-mode of the scalar field is motivated by the full sphaleron
solution of the earlier sections, which only turns on the lowest
spherical vector harmonics.  Naturally, in the full U(4) there are
also operators of the form $\Tr(D_\mu\phi D_\nu\phi)$. However, in the
oscillator picture these are turned on by the oscillators that create
the higher spherical tensorial harmonics.}  The proper linear
combinations of these operators are
\begin{equation}
\hat{O}_2^\dagger = \Tr( a^\dagger a^\dagger )\,,\qquad
\hat{O}_4^\dagger = \Tr( a^\dagger a^\dagger a^\dagger a^\dagger ) -
\frac{2N^2 + 1}{N(N^2+2)} \Tr( a^\dagger a^\dagger ) \Tr( a^\dagger
a^\dagger )\,.
\end{equation}
These lead to $\langle \hat{O}_4\,|\, \hat{O}_2
\hat{O}_2\rangle=0$. Multi-particle states will generically not be
orthogonal, but in our case this turns out to be far less important
than the $1/N^2$~corrections to the norms. We will for simplicity also
use a classical configuration for which
\begin{equation}
\frac{\eta_4}{N} = \left(\frac{\eta_2}{N}\right)^2 = \frac{\eta}{N}\,,
\end{equation}
where the $\eta_J$ are defined in~\eqn{argu}. Closer inspection of the
coherent state of the sphaleron given in~\eqn{coherent} shows that the
expectation values of e.g.~the $\Tr(F_{mn} F^{mn})$ and~$\Tr(F_{m
n}F^{m n} F_{rs} F^{rs})$ states are similarly related.

\begin{figure}[t]
\begin{center}
\includegraphics*[width=.6\textwidth]{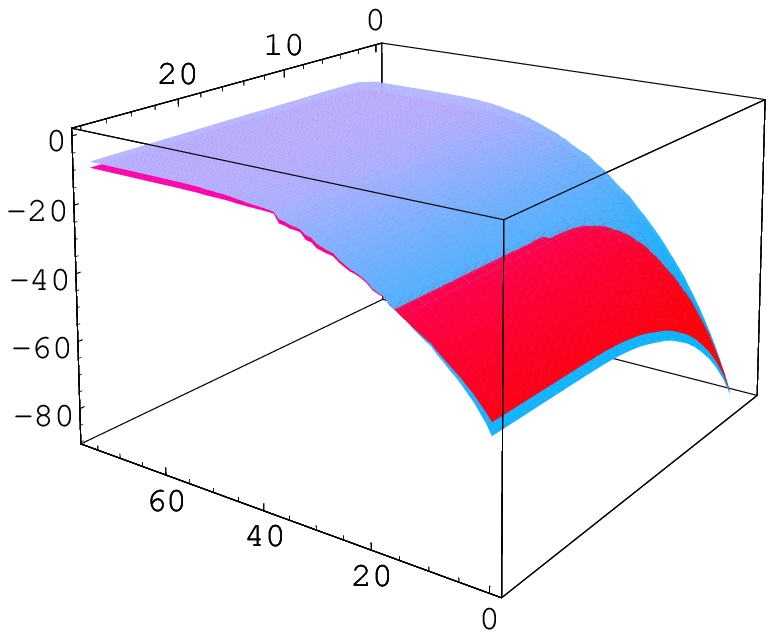}
\makebox[0pt]{\hspace{-1.3cm}\raisebox{2.2cm}{\includegraphics*[height=5cm]{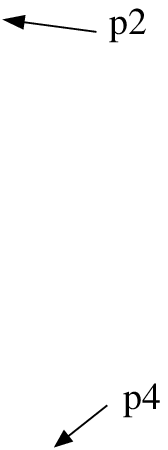}}}
\makebox[0pt]{\hspace{-19cm}\raisebox{5.4cm}{\includegraphics*[width=2cm]{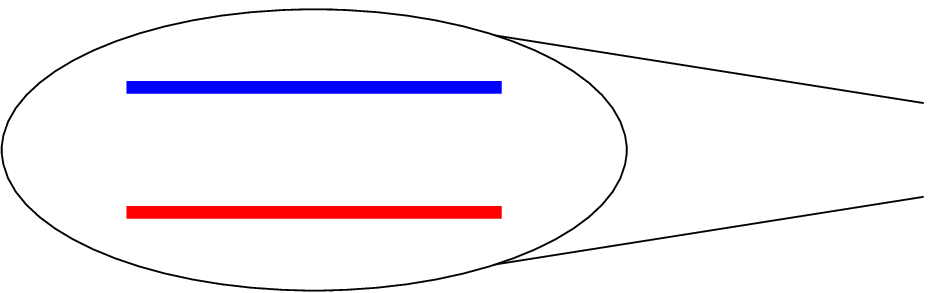}}}
\caption{Successive approximations to the logarithm of the total
energy radiated in the $J=2$ particles (light, blue surface) and $J=4$
particles (dark, red surface). The $x$ and $y$ axes label the maximum
value of $p_2$ and $p_4$ in the sum~\eqn{e:cutoffsum}. The values
asymptote to the full result in the upper left corner of the
graph. While the present plot shows energies, qualitatively similar
plots are obtained for the particle numbers.}
\label{f:energies}
\end{center}
\end{figure}

The energy radiated into $O_{J=2}$ and $O_{J=4}$ particles can be computed
using formula~\eqn{e:energies}, summed over a suitably large range of
values for $p_2$ and $p_4$. In our particular case, this formula
reduces to
\begin{equation}
\label{e:cutoffsum}
E(J,p^{\text{\maxp}}_2, p^{\text{\maxp}}_4) =
  \sum_{p_2=0}^{p_2^{\text{\maxp}}} \sum_{p_4=0}^{p_4^\text{\maxp}}
  \left|\frac{\eta_2^2}{\lambda^2}\right|^{p_2}
  \left|\frac{\eta_4^2}{\lambda^4}\right|^{p_4}
  \frac{J p_J}{2^{p_2} 4^{p_4}} 
  \frac{{\mathcal C}^2}{\langle 0|\, (\hat{O}_2)^{p_2} (\hat{O}_4)^{p_4}\,
                  (\hat{O}_4^\dagger)^{p_4} (\hat{O}_2^\dagger)^{p_2}\,
                  |0\rangle\,\langle c|c\rangle}\,.
\end{equation}
and the maximum values of $p_2$ and $p_4$ which are included in the
sum should be taken sufficiently large as to include at least the
maximum term in the sum. This requirement is indeed met in our
numerical approach.  We have computed the ratio of energies in the
$J=2$ and $J=4$ particles using successive approximations
of~\eqn{e:cutoffsum}, for larger and larger $p^{\text{\maxp}}_2$ and
$p_4^{\text{\maxp}}$, for a range of couplings.\footnote{From the
gravity point of view, in case of large $N$, the $p^{\text{\maxp}}$
should always be such that the total energy (i.e.~conformal dimension)
carried by this multi particle state is smaller than~$N^2$ in order to
neglect back reaction. In the case of small~$N$, such as discussed
here, constants of order one become relevant, and this rough estimate
is no longer sufficient. For example, it turns
out~\cite{Peeters:2004rd} that the maximal probability for the number
of particles of type ${\mathcal O}_2$ is larger than $N^2$, but the
total energy carried by these particles is still smaller than the
energy of the brane (once constants of order one have been taken into
account).} A typical example is plotted in
figure~\ref{f:energies}. One clearly sees that the asymptotic value of
the ratio $E(4)/E(2)$, given by the exponent of the asymptotic height
difference between the two surfaces, is smaller than one. We therefore
conclude that our calculation predicts that higher-energy states in
the decay product are suppressed with respect to the lower-energy
ones. This is in qualitative agreement with alternative calculations
of this decay process~\cite{Lambert:2003zr}.

It would be very interesting to extend our analysis to higher-rank
gauge groups, perhaps by obtaining an analytic expression for the
norms of the states. For~$N>4$, there are more than two gauge singlet
states, and it becomes possible to determine the suppression factor as
a function of the energy in more detail. We leave this for future
investigations.
\psfrag{10}{$\scriptstyle 10$}
\psfrag{20}{$\scriptstyle 20$}
\psfrag{40}{$\scriptstyle 40$}
\psfrag{60}{$\scriptstyle 60$}
\psfrag{-20}{$\scriptstyle ~-20$}
\psfrag{-40}{$\scriptstyle ~-40$}
\psfrag{-60}{$\scriptstyle ~-60$}
\psfrag{-80}{$\scriptstyle ~-80$}
\psfrag{0}{$\scriptstyle 0$}
\psfrag{p2}{$\scriptstyle p_2^\text{\maxp}$}
\psfrag{p4}{$\scriptstyle p_4^\text{\maxp}$}

\section{Summary and outlook}

We have presented the formalism to analyse the decay of unstable
D-branes in the~$\text{AdS}_5 \times S^5$ background by considering
the dual gauge theory.\footnote{Such dynamical features of the
correspondence have meanwhile also been studied in the context of
large spinning strings~\cite{Peeters:2004pt}.} Our results show
qualitative agreement with previous work on D-particle decay, and our
work provides a basis for further study of non-perturbative dynamical
features of the correspondence.

A relevant way of improving on our results would be to determine
analytical expressions for the norms required in
section~\ref{s:U4results} (using the construction of states in terms
of group
characters~\cite{Balasubramanian:2001nh,Corley:2001zk,Kristjansen:2002bb}). This
would allow one to extend the results obtained there to large values
of~$N$.  Also, as we have explained, due to the non-perturbative
nature of the initial sphaleron configuration, the computation of the
decay product requires information from a regime in which
both~\mbox{$N\rightarrow\infty$} as well as the number of particles
~$p\rightarrow\infty$. Knowing the norms of states analytically should
allow us to understand this double limit. This may perhaps circumvent
the need to calculate the norms of states exactly when calculating the
energy distribution in the final state. Finally, it would be
interesting to understand how quantum corrections can be incorporated
into our formalism, in order to see how much they influence the
qualitative characteristics of the decay product.

\section*{Acknowledgements}

We would like to thank Gleb Arutyunov, Rajesh Gopakumar, Justin
David, Stefano Kovacs, Charlotte Kristjansen, Shiraz Minwalla, Jan
Plefka and Ashoke Sen for discussions.
\vspace{-10ex}

\begingroup\raggedright\endgroup

\end{document}